\documentclass[12pt]{article}

\usepackage{epsfig,amsmath,amssymb,graphics,verbatim,amsfonts,subfigure,psfrag}
\usepackage[letterpaper,margin=0.85in]{geometry}
\usepackage{xcolor}
\usepackage{amsthm}
\usepackage{enumitem}
\usepackage{overpic} 
\usepackage{hyperref}

\def\Id{{\textnormal{Id}}}

\def\ra{\rightarrow}
\def\I{\infty}

\def\R{\mathbb{R}}

\def\N{\mathbb{N}}
\def\E{\mathbb{E}}
\def\I{\infty}

\newcommand{\be}{\begin{equation}}
\newcommand{\ee}{\end{equation}}
\newcommand{\bea}{\begin{eqnarray}}
\newcommand{\eea}{\end{eqnarray}}
\newcommand{\beann}{\begin{eqnarray*}}
\newcommand{\eeann}{\end{eqnarray*}}
\newcommand{\benn}{\begin{equation*}}
\newcommand{\eenn}{\end{equation*}}

\def\ra{\rightarrow}
\def\I{\infty}

\newcommand{\cC}{{\mathcal C}}  
\newcommand{\cH}{{\mathcal H}}  
\newcommand{\cL}{{\mathcal L}}  
\newcommand{\cN}{{\mathcal N}}  
\newcommand{\cO}{{\mathcal O}}  
\newcommand{\cP}{{\mathcal P}}  
\newcommand{\cX}{{\mathcal X}}  

\def\txta{{\textnormal{a}}}

\def\txtd{{\textnormal{d}}}
\def\txte{{\textnormal{e}}}

\def\txtr{{\textnormal{r}}}

\def\txtD{{\textnormal{D}}}

\begin{document}

\title{A Note on Kernel Methods for Multiscale Systems with Critical Transitions}

\author{Boumediene Hamzi\\Department of Mathematics, AlFaisal University, Riyadh/Saudi Arabia \\ email: {boumediene.hamzi@gmail.com} \and Christian Kuehn\\ Faculty of Mathematics, Technical University of Munich, Munich/Germany  \\ email: {ckuehn@ma.tum.de} \and Sameh Mohamed\\ Singapore University of Technology and Design (SUTD), Singapore  \\ email: {sameh@sutd.edu.sg}}

\date{}
\maketitle

\begin{abstract}
We study the maximum mean discrepancy (MMD) in the context of critical transitions
modelled by fast-slow stochastic dynamical systems. We establish a new link between
the dynamical theory of critical transitions with the statistical aspects of the MMD.
In particular, we show that a formal approximation of the MMD near fast subsystem 
bifurcation points can be computed to leading-order. In particular, this leading
order approximation shows that the MMD depends intricately on the fast-slow
systems parameters and one can only expect to extract warning signs under rather 
stringent conditions. However, the MMD turns out to be an
excellent binary classifier to detect the change point induced by the critical 
transition. We cross-validate our results by numerical simulations for a van 
der Pol-type model.
\end{abstract}


\section{Introduction}

Drastic sudden large events in dynamical systems have become a key area of interest
in a broad range of applications~\cite{AshwinWieczorekVitoloCox,Schefferetal}. From 
the perspective of 
modelling, a successful framework to capture many critical transitions has been to use 
systems with multiple time scales in combination with bifurcation theory~\cite{Kuznetsov}.
The idea is that there are fast variables, which are driven slowly towards a bifurcation
point, where the system can undergo a sudden jump for certain types of bifurcations. 
One aim in this context is to determine, whether there are early-warning signs for critical
transitions, which can be computed from time series data before the actual event occurred.
Ground-breaking work by Wiesenfeld in the 1980s~\cite{Wiesenfeld1} has already clearly
shown that pre-cursors of bifurcations exist, and that they can be extracted from
stochastic fluctuations based upon critical slowing down. For example, for some class 
of systems, a key insight is 
that changes in statistics, such as growing variance and autocorrelation, can be used as a 
basis for early-warning signs. This idea has recently been
(re-)discovered in various applications, e.g.~in ecology~\cite{CarpenterBrock}. Mathematically
there is a key interplay between small noise levels, a small time-scale separation parameter
and the proximity to a bifurcation instability~\cite{KuehnCurse} but one can indeed build
a surprisingly detailed theory for this setting in the context of fast-slow stochastic 
ordinary differential equations (SODEs)~\cite{KuehnCT2}; see also~\cite{BerglundGentz,KuehnBook}
for more detailed technical background as well as Section~\ref{sec:CT}. However, in advance 
of asking for prediction, we may also ask for detailed methods to distinguish critical transitions
and/or extreme events in dynamical systems from regular slow/gradual transitions. Hence, we
are faced with the following challenges:

\begin{itemize}
 \item \emph{Change-point detection:} In a time series generated by a fast-slow
SODE, there could be many different types and sizes of drastic jumps. Hence,
it would not only be useful to have an automatic and generic 
classifiers~\cite{Fryzlewicz,Liuetal,SiegemundVenkatraman}, when we actually 
observe a critical transition but also to cross-validate a classifier against 
explicit low-dimensional models.
 \item \emph{Machine learning:} One would like to be able to link 
learning algorithms to the theory of critical transitions to gain a better
understanding of learning techniques via concrete dynamical systems. 
Although prediction of dynamics via machine learning 
is frequently applied in practice~\cite{Wittenetal} and preliminary results 
were obtained for seizure detection \cite{Hamzi2}, a fully-proven theory 
for this approach for nonlinear systems with critical transitions does not 
exist. Hence, linking techniques from this area
to low-dimensional tractable mathematical models should turn out to be very 
useful.
 \item \emph{Data reliability:} It is evident that with an extremely sparse 
data set, we are not going to be able to understand and/or predict critical
transitions. Yet, with perfectly sampled data, the theory is proven to work.
To determine the precise boundary between these two regimes is 
crucial~\cite{BoettingerHastings,DitlevsenJohnsen,ZhangHallerbergKuehn}. Therefore,
a better link to statistical methods is necessary.
\end{itemize}

All three problems seem to strongly suggest to try to merge and complement
the theory of critical transitions with (kernel-based) methods originating
from functional analysis and frequently used in machine learning and statistics. Here we 
take the \emph{first very small steps to achieve the link between kernel methods 
and fast-slow SODEs} in the context of critical 
transitions. Our main tool on the statistical and machine learning side is
the maximum mean discrepancy (MMD)~\cite{Borgwardtetal,Grettonetal,
sri}. The MMD essentially provides a measure for the 
difference between two probability distributions. Of course, one can also
first estimate two densities and then compare them with suitable metrics
(e.g., total variation, Wasserstein, Kullback-Leibler, etc). Yet, density
estimators are expensive, have computational limitations, and it can be
difficult to theoretically link the two-stage process of estimation and 
comparison to concrete dynamical systems. The MMD provides a more direct approach 
and it is relatively easy to compute, so it is one natural starting point for 
our goal to link the - currently quite distant - areas we study here.

The main idea we present in this work is to use the standard fast-slow SODEs
as an explicit test scenario for the MMD. In particular, we formally use a
quasi-steady approximation of the local distribution as a sample path approaches
a critical transition point. Using this technique, we can provide an 
approximation of the MMD and more explicitly understand its dependency upon
the parameters and upon the current state in phase space of the dynamics.
This establishes a concrete link between a crucial model class and kernel-based
statistical methods. More precisely, we show that the MMD is 
excellent in classifying critical transitions in fast-slow SODEs as change
points, yet, since it is kernel dependent and depends crucially on the sliding window
choice, it can be a lot more difficult to extract a practical early-warning 
sign from it. As a final step, we also cross-validate our
results with numerical simulations of a van der Pol-type model.

The paper is structured as follows: Sections~\ref{sec:CT}-\ref{sec:kernel} collect
the relevant tools we need from fast-slow SODEs as well as from kernel-based 
methods and the MMD; we have included this background also in the anticipation
of the different and currently very disjoint communities, which we aim to bring 
closer together. Section~\ref{sec:OU} contains the new connection between
the two areas and our formal approximations of the MMD near slow manifolds
in SODEs. In Section~\ref{sec:models}, we provide the numerical simulations for
a van der Pol-type model, which confirm our formal analytical calculations.   

\section{Critical Transitions in Multiple Time Scale Systems}
\label{sec:CT}

In this section, we provide a brief review of critical transitions theory and
associated dynamical systems models as this theory might not be known to experts
in machine learning and kernel methods.
One flexible and broad class of models for critical transitions are given by
fast-slow stochastic ordinary differential equations (SODEs) of the form
\be
\label{eq:fs}
\begin{array}{lcr}
\txtd x &=& \frac1\varepsilon f(x,y)~\txtd s+\frac{\sigma_f}{\sqrt\varepsilon}
F(x,y)~\txtd W,\\ 
\txtd y &=& g(x,y)~\txtd s+\sigma_g G(x,y)~\txtd W,\\ 
\end{array}
\ee
where the maps $f:\R^{m+n}\ra \R^m$, $g:\R^{m+n}\ra \R^n$, $F:\R^{m+n}\ra 
\R^{n\times k}$, $G:\R^{m+n}\ra \R^{m\times k}$ are assumed to be sufficiently
smooth, $W=W(s)\in\R^k$ is a vector of $k$ independent Brownian motions, 
$0<\varepsilon\ll 1$ controls the time scale separation between the fast
variables $x=x(s)\in\R^m$ and the slow variables $y=y(s)\in\R^n$, while 
$\sigma_f,\sigma_g>0$ control the noise levels; we also set $\sigma_*^2:=
\sigma_g^2+\sigma_f^2$ and consider a given deterministic initial condition 
$(x(0),y(0))$. One frequently assumes that $\sigma_f=
\sigma_f(\varepsilon)$ and $\sigma_g=\sigma_g(\varepsilon)$ are small functions 
of $\varepsilon$ and also vanish in the asymptotic limit $\varepsilon\ra 0$.
The main idea to model critical transitions using~\eqref{eq:fs} is to interpret the 
slow variable(s) $y$ as drifting parameters, which move the fast variables $x$
towards a point, where the system can undergo a large transition within a short time. 
More precisely, consider first the deterministic case $\sigma_*=0$.
This yields the fast-slow ordinary differential equation (ODE)
\be
\label{eq:fs1}
\begin{array}{rcrcr}
\varepsilon\frac{\txtd x}{\txtd s}&=&\dot{x}&=& f(x,y),\\ 
\frac{\txtd y}{\txtd s} &=& \dot{y} &=& g(x,y).\\ 
\end{array}
\ee
The set $\cC_0:=\{(x,y)\in\R^{m+n}:f(x,y)=0\}$ is called the critical set. We shall
assume it is a manifold here and refer to it as the critical manifold. $\cC_0$ is
called normally hyperbolic at $p\in\cC_0$ if the matrix $\txtD_x f(p)\in\R^{m\times 
m}$ has no eigenvalues with zero real parts. In this case, Fenichel's 
Theorem~\cite{Fenichel4,Jones,KuehnBook} provides a (regular) perturbation theory
for the critical manifold $\cC_0$ to a slow manifold $\cC_\varepsilon$. Locally,
we also have that 
\benn
\cC_0=\{x=h_0(y)\},\qquad \cC_\varepsilon=\{x=h(y)=h_0(y)+\cO(\varepsilon)\}
\eenn 
so we may present $\cC_0$ as a graph and the fast variables $x$ are slaved to
the slow variables $y$. Dynamically,
these results imply that the dynamics for $0< \varepsilon \ll1$ near 
$\cC_\varepsilon$ is well-approximated by the slow flow of the differential
algebraic equation
\be
\label{eq:fs2}
\begin{array}{rcr}
0 &=& f(x,y),\\ 
\dot{y} &=& g(x,y),\\ 
\end{array}
\ee
which is also known as the slow subsystem. Furthermore, the fast dynamics near
$\cC_\varepsilon$ is well-approximated by the fast subsystem 
\be
\label{eq:fs3}
\begin{array}{rcrcl}
\frac{\txtd x}{\txtd t} &=& x' &=& f(x,y),\\ 
\frac{\txtd y}{\txtd t} &=& y' &=& 0,\\ 
\end{array}
\ee
obtained from~\eqref{eq:fs1} by changing from the slow time $s$ to the fast 
time $t:=s/\varepsilon$ and then taking the limit $\varepsilon\ra 0$. 
In~\eqref{eq:fs3}, the $y$ variables are just parameters. A 
fast subsystem bifurcation of an equilibrium point $p$ with $f(p)=0$ just means 
that normal hyperbolicity is lost at $p_*=(x_*,y_*)$. At these bifurcation points, 
the fast and slow scales interact in a non-trivial way~\cite{KuehnBook}, which can
lead to large jumps in trajectories of the full system~\eqref{eq:fs1} for
certain types of bifurcations. A classification of these (bifurcation-induced)
critical transitions is given in~\cite{KuehnCT2}. As a simple example consider
the normal form for a fast-slow fold singularity 
\be
\label{eq:fold}
\begin{array}{rcl}
 x' &=&y-x^2,\\ 
 y' &=& -\varepsilon,\\ 
\end{array}
\ee 
where $\cC_0=\{y=x^2\}$ splits into two parts $\cC_0^\txta=\cC_0\cap 
\{x>0\}=\{x=h_0^\txta(y)=\sqrt{y}:y>0\}$ and $\cC_0^\txtr=\cC_0\cap \{x<0\}=
\{x=h_0^\txtr(y)=-\sqrt{y}:y>0\}$ and the fold singularity/point at the 
origin $0:=(0,0)=p_*$. Since $\txtD_x f(x,y)=-2x$, we conclude that $\cC^\txta_0$
(resp.~$\cC^\txtr_0$) is normally hyperbolic and attracting (resp.~repelling), when viewed
as equilibrium points for~\eqref{eq:fs3}. It is easy to check using Fenichel's
Theorem that if we start near $\cC_0^\txta$ for $y(0)>0$, $y=\cO(1)$ as 
$\varepsilon\ra 0$, $y(0)\approx x(0)^2$ then we get attracted to a slow
manifold $\cC_\varepsilon^\txta$, which is located at a distance 
$\cO(\varepsilon)$ from $\cC_0$. The slow dynamics on $\cC_\varepsilon^\txta$
is governed by $\dot{y}=-1$ so that we slowly approach a neighbourhood of the 
fold point in finite time. In a neighbourhood of the fold point, one can 
prove that a jump occurs directed towards $x=-\I$; even the precise local
asymptotics in $\varepsilon$ is 
well-known~\cite{DumortierRoussarie,KruSzm3,KuehnBook,MisRoz}. 
The main point  for the work here is that a particular class of critical transitions 
occurs in fast-slow systems and that suitable simplifications
in these systems to provide analytical guidelines for time series analysis.

Next, we briefly clarify the role of noise ($\sigma_*\neq 0$). In actual time 
series measurements of critical transitions, we always have to
deal with noise. This may either originate from internal system fluctuations,
external forcing and/or measurement errors. Hence, we include
stochasticity directly into the modelling framework. Furthermore, it turns out
that noise can be useful in the prediction of critical transitions.
The idea is that fluctuations can help to detect critical slowing 
down~\cite{Wiesenfeld1}, i.e., the linearization $\txtD_xf(p)=:A(p)$ is close 
to being singular near the critical transitions at $p_*$. Indeed, if we consider 
the linearized fast-subsystem SODE around an attracting branch of 
$\cC_0$~\cite{BerglundGentz,KuehnCT2} for the simplest additive noise 
($F\equiv \Id$, $k=m$, $\sigma_g=0$) we get 
\be
\txtd X=\txtD_x f(x,y)X~\txtd t +\sigma_f~\txtd W=:A(p)X~\txtd t+\sigma_*~\txtd W, 
\ee  
which is just an Ornstein-Uhlenbeck (OU) process $X=X(t)\in\R^m$. As long as 
$A(p)$ has spectrum in the left-half of the complex plane, one easily checks
that the covariance matrix $V(t):=\textnormal{Cov}(X(t))$ has a stationary limit 
\benn
V_\I=\lim_{t\ra +\I}V(t).
\eenn
However, in the limit $p\ra p_*$, there are elements in the covariance matrix,
which diverge; see~\cite{KuehnCT2} for more details. For this work, it is only
relevant to know that - in practical terms - we expect the (co-)variance to
grow near a critical transition modeled by a sufficiently generic fast-slow 
system. Of course, when the fast jump occurs, the (quasi-stationary approximation 
of the) mean of the fast variable stochastic process
changes drastically, when compared to the mean during the slow phase approaching the
fast subsystem bifurcation.

\section{Kernels and Maximum Mean Discrepancy}
\label{sec:kernel}

In this section, we give a brief review of reproducing kernel Hilbert 
spaces as used in statistical learning theory as this theory might not be
known to experts in critical transitions. The discussion here mostly
follows~\cite{CuckerSmale,Wahba,SchoelkopfSmola}; for a historical 
perspective see also~\cite{Schoenberg,Aronszajn} and for a recent
application related to dynamics see~\cite{Hamzi1}. Let $\cH$ be a 
Hilbert space of functions on a separable metric space $\cX$; more 
generally, many constructions work for just a set $\cX$ but we shall not
explore this direction here. Denote by 
$\langle u, v \rangle$ the inner product on $\cH$ for $u, v \in \cH$ 
and by $\|u\|= \langle u, u \rangle^{1/2}$
the induced norm. $\cH$ is called a reproducing kernel Hilbert space 
(RKHS) if there exists a kernel 
\benn
K:\cX \times \cX \rightarrow \R
\eenn
such that $K$ has the reproducing property, i.e., $\forall u \in \cH$, 
$u(x)=\langle u(\cdot),K(\cdot,x) \rangle$ and $K$ spans $\cH$, i.e., 
$\cH=\overline{\textnormal{span}\{K(x,\cdot)|x \in \cX\}}$. $K$ will be called 
a reproducing kernel of $\cH$. $\cH_K(\cX)$ will denote the RKHS $\cH$ with 
reproducing kernel $K$. The idea to consider kernels naturally builds upon
finite-dimensional matrix analysis. A function $K$ is called a Mercer kernel 
if it is continuous, symmetric and positive definite. The following kernels, 
defined on a compact domain $\cX \subset \R^n$, are Mercer kernels:
\begin{itemize}
 \item[(M1)] linear: $K(x,z)=\kappa x^\top z$ for $\kappa>0$;
 \item[(M2)] polynomial: $K(x,z)=(1+x^\top z)^p$ for $p \in \N$; 
 \item[(M3)] Gaussian: $K(x,z)=\exp\left(-(x-z)^\top(x-z)/\sigma^2\right)$ for $\sigma >0$.
\end{itemize}
If $K$ is a Mercer kernel on $\cX$, then there is a unique Hilbert space $\cH$ 
of functions on $\cX$ so that $K$ is a reproducing kernel. Moreover, kernels can
be viewed as generalized dot products. There exists a feature map $\Phi: \cX 
\rightarrow \cH$ such that
\benn
K(x,z)= \langle \Phi(x), \Phi(z) \rangle_\cH \quad \mbox{for} \quad x,z \in \cX.
\eenn
The dimension of the RKHS can be infinite and corresponds to the dimension of the 
eigenspace of the integral operator $\cL_K: L_{\nu}^2(\cX) \rightarrow C(\cX)$ defined 
\benn
(\cL_K u)(x)=\int K(x,z)u(z)~\txtd\nu(z),\qquad u \in L_{\nu}^2(\cX), 
\eenn 
where $\nu$ is a Borel measure on $\cX$. Conversely, for a given $\cH$ of functions 
$u: \cX \rightarrow \R$, with $\cX$ compact, satisfying
\benn
\forall x \in \cX, \exists \kappa_x>0, \quad \mbox{such that} \quad |u(x)| \le 
\kappa_x \|u\|_\cH,
\eenn
one can prove that $\cH$ has a reproducing kernel $K$. In practice, one frequently 
just takes the canoncial feature map $\Phi(x)=K(x,\cdot)$ for a chosen Mercer kernel
such as (M1)-(M3). 

RKHSs play an important role in change-point detection. Suppose we are given a sequence of samples 
$x_1$, $x_2,\cdots,x_M$ from a domain $\cX$. We want to detect a possible change-point 
time $\tau$, such that before $\tau$, the samples $x_i \sim P$ i.i.d for $i \le \tau$, 
where $P$ is the so-called background distribution, and after the change-point, the 
samples $x_i \sim Q$ i.i.d for $i \ge \tau+1$, where $Q$ is a post-change distribution.
Hence, the problem is to compare two distributions $P$ and $Q$. Change-point 
detection in RKHSs is based on mapping the dataset into a RKHS $\cH$ and to 
compute a measure of heterogeneity, which is small if $P=Q$ and large if $P$ 
and $Q$ are far apart. Several measures of heterogeneity have been proposed in 
the literature; cf.~\cite{Harchaouietal} for a survey. We are going to use the 
maximum mean discrepancy (MMD)  
\be
\label{eq:MMDdef1}
\mbox{MMD}[\cH, P, Q] := \sup_{u \in \cH,\|u\|\leq 1}
\{\mathbb{E}_x[u(x)]-\mathbb{E}_z[u(z)] \},
\ee
as a measure of heterogeneity. We also note an alternative way to express the 
MMD~\cite{Grettonetal}. Let $\Phi: \cX \rightarrow \cH$ be a continuous feature 
mapping and assume that $K$ is measurable and bounded, i.e., $\sup_{x\in\cX} 
K(x,x) < \infty$. Let $\cP$ be the set of Borel probability measures on $\cX$. 
We define the mapping to $\cH$ of $P \in \cP$ as the expectation of $\Phi(x)$ 
with respect to P
$$
\begin{array}{rcl} 
\mu_P: \cP& \rightarrow & \cH  \\
P & \mapsto& \int_\cX \Phi(x)~\txtd P
\end{array}
$$
The maximum mean discrepancy (MMD) between two probability measures is then given 
as the distance between two such mappings~\cite{Grettonetal}
\bea
\mbox{MMD}[\cH,P,Q] &=& \|\mu_P-\mu_Q\|_\cH, \label{eq:MMDdef2}\\
&=& \big(\mathbb{E} _{x,{x}^{\prime}}[K(x,x^{\prime})]+\mathbb{E} _{y,{y}^{\prime}}
[K(z,z^{\prime})]-2 \mathbb{E} _{x,z}[K(x,z]\big)^{\frac{1}{2}}\nonumber
\eea
where $x$ and $x^{\prime}$ are independent random variables drawn according to $P$, 
$z$ and $z^{\prime}$ are independent random variables drawn according to $Q$, and $x$ 
is independent of $z$. This quantity is a pseudo-metric on distributions\footnote{Given a set $M$, a metric for $M$ is a function $\rho:M \times M \rightarrow \mathbb{R}^+$  such that i.) $\forall x \in M, \rho(x,x)=0$, ii.) $\forall x,y \in M, \rho(x,y)= \rho(y,x)$, iii.) $\forall x,y,z \in M, \rho(x,z) \le \rho(x,y)+\rho(y,z)$, and iv.) $\rho(x,y) = 0 \implies x = y$.  A pseudometric only satisfies (i)-(iii) of the properties of a metric. Unlike a metric space $(M,\rho)$, points in a pseudometric space need not be distinguishable and one may have $\rho(x, y) = 0$ for $x \ne y$.}. For the MMD 
to be a metric, it is sufficient that the kernel is characteristic, i.e., the map 
$\mu_P: \cP \rightarrow \cH $ is injective. This is satisfied by the Gaussian kernel 
(both on compact domains and on $\R^d$) for example \cite{sri}. The formulation~\eqref{eq:MMDdef1}
of the MMD will be more useful for theoretical considerations as discussed in 
Section~\ref{sec:OU} while~\eqref{eq:MMDdef2} is going to be more convenient for
the estimation of the MMD from time series as discussed in Section~\ref{sec:models}. 

\section{Linearization and Maximization}
\label{sec:OU}

In this section, we would like to show explicitly that for bifurcation-induced 
critical transitions in fast-slow SODEs~\eqref{eq:fs}, the MMD indeed detects the 
corresponding change points. Our calculations in this work are \emph{formal}, i.e.,
we use several \emph{approximations} without rigorous proof. The idea here is to
develop the correct intuition and then study the situation also numerically in
Section~\ref{sec:models}. Detailed proofs are beyond the scope of this work and 
will be considered in a future study. 

We consider the two-dimensional setting $m=1=n$ 
for~\eqref{eq:fs} as it contains already all the main features. Furthermore, the 
definition(s)~\eqref{eq:MMDdef1}-\eqref{eq:MMDdef2} implicitly or explicitly 
involve the kernel $K$. Therefore, we would like to consider several basic classes
of kernels and study the parameter dependence of the MMD in these situations near
a critical transition. Let $(x_*(t),y_*(t))\in\R^2$ be a trajectory of a deterministic
fast-slow ODE (with $m+n=2$) lying in an attracting part of the critical manifold 
$\cC_0^\txta$ such that this trajectory reaches at some time $T_*>0$ a bifurcation point 
of the fast subsystem, which is a critical transition~\cite{KuehnCT2}. Then it is a 
standard technique to linearize the fast variable part of the SODE around the 
deterministic reference solution to obtain 
\be
\label{eq:OU1D}
\txtd X = [\partial_x f(x_*(t),y_*(t))](X+b(t))~\txtd t+\sigma_f~\txtd W
=:(-a(t)X+a(t)b(t))~\txtd t+\sigma_*~\txtd W,
\ee
in the case of additive noise; under suitable assumptions on $\sigma_g$ and $G$, the slow
variable noise terms only enter as higher-order corrections in~\eqref{eq:OU1D} so we 
shall drop them here. Classically one would also aim to remove the mean $b(t)$ from
\eqref{eq:OU1D} if one only wants to study the variance~\cite{BerglundGentz,KuehnCT2}
but here it is crucial to keep it as a varying parameter. The 
SODE~\eqref{eq:OU1D} is a \emph{time-dependent} OU process 
providing a good lowest-order approximation $X=X(t)\in\R$ to the local fluctuations 
transverse to $\cC_\varepsilon$ of the fast variable $x$; 
see~\cite{BerglundGentz,KuehnCT2,KuehnBook}
for a more detailed derivation and the validity range of the approximation. At the 
critical transition point, we lose normal hyperbolicity so that 
\benn
a(t)>0\quad \text{for $t\in[0,T_*)$} \qquad \text{and}\qquad a(T_*)= 0,
\eenn
where the positive sign of $a(t)$ arises because we introduced an additional negative
sign in the definition~\eqref{eq:OU1D} as it simplifies the notation below. 
Note that $a(t)$ is essentially only dependent upon $y_*(t)$ as we locally can 
always write $x_*(t)=h_0^\txta(y_*(t))$. The dynamics of $y_*$ is given by the slow 
flow~\eqref{eq:fs2}. In particular, $y_*(t)$ is only slowly varying so the same holds
for $a(t)$. If we measure the original process $x(t)$ frequently enough over a given
time interval, we can make a quasi-steady assumption that within a time window
$t\in[t_j,t_{j+1}]$ we approximately have 
\bea
\label{eq:OUstat}
\txtd X &=& (-a((t_j+t_{j+1})/2)X+a((t_j+t_{j+1})/2)b((t_j+t_{j+1})/2))~\txtd t
+\sigma_*~\txtd W \nonumber \\ &=:&b_ja_j-a_jX~\txtd t+\sigma_*~\txtd W,
\eea
where $a_j>0$ for $t_{j+1}\leq T_*$. It is well-known~\cite{Gardiner} that the OU 
process defined by~\eqref{eq:OUstat} has a stationary distribution with Gaussian 
density
\be
\label{eq:OUrho}
\rho(x;a_j,b_j,\sigma_*) = \sqrt{\frac{a_j}{\pi\sigma_*^2}}
\txte^{-a_j(x-b_j)^2/\sigma_*^2}.
\ee
It is possible to check that as long as we keep $\sigma_*>0$ fixed, then the variance
of $X(t)$ diverges near a fast subsystem bifurcation point as 
$t\nearrow T_*$~\cite{KuehnCT2}. Furthermore,
we can always re-scale $a_j$ using the definition $\alpha_j:=a_j/\sigma_*^2$ in the
stationary density to eliminate one parameter 
\be
\label{eq:OUrho1}
\rho(x;\alpha_j,b_j) = \sqrt{\frac{\alpha_j}{\pi}}\txte^{-\alpha_j(x-b_j)^2}.
\ee
In principle, the density $\rho$ has support on $\R$, yet it is well-established
for critical transitions that we are only interested in the mean and the local 
fluctuations on time scales before a noise-induced large deviation effect has taken 
place. Therefore, we are going to approximate the process on a large compact interval
$\cX$ containing $b(t)$. Now we can finally start to study, how a kernel-based method 
detects the change point, and how it interacts with critical transitions in fast-slow 
SODEs. We can use the MMD to compare two stationary densities
\be
\label{eq:MMDdefuseit}
\mbox{MMD}[\cH,P_i, P_j] := \sup_{\|u\|\leq 1}
\{\mathbb{E}_x[u(x)]-\mathbb{E}_y[u(y)] \},
\ee
where $u\in\cH$, $x\sim P_i$ has the density $\rho(x;\alpha_i,b_i)$ and $y\sim P_j$ 
has the density $\rho(y;\alpha_j,b_j)$. Recall that 
\benn
\E_x[u(x)]=\E_x[\langle u,K(x,\cdot)\rangle]=\langle u,\E_x[K(x,\cdot)]\rangle.  
\eenn
Selecting the linear kernel $K(x,z)=\kappa xz$ from (M1) we get
\be
\label{eq:lininsert}
\E_x[K(x,z)] = z\kappa\int_\cX x \sqrt{\frac{\alpha_i}{\pi}}\txte^{-\alpha_i(x-b_i)^2}~\txtd x
\approx z\kappa\int_\R x \sqrt{\frac{\alpha_i}{\pi}}\txte^{-\alpha_i(x-b_i)^2}~\txtd x=
\kappa z b_i
\ee
so we just extract the mean. The MMD is then given by
\benn
\mbox{MMD}_{\textnormal{(M1)}}[\cH,P_i, P_j] \approx \kappa (b_i-b_j) \sup_{\|u\|\leq 1}
\langle u,\Id \rangle,\qquad \Id(z):=z.
\eenn
The last supremum obviously depends upon $\cH$ but does not depend upon any parameters
of the fast-slow SODE. In particular, for the linear kernel, only the variation in the
mean is picked up. This is expect since the linear kernel is only characteristic in detecting
the mean, i.e., for a general polynomial kernel with power $p$, the MMD returns zero
if the first $p$ moment coincide. However, even for the linear kernel and for some critical 
transitions, this can be sufficient for detecting them. For example, in the case of a 
fold~\eqref{eq:fold}, we easily get
\benn
b(t)=\sqrt{-\varepsilon t+y(0)}
\eenn
so the difference between $b_i$ and $b_j$ increases as $t\ra T_*$. However, for other 
bifurcations such as the pitchfork with $f(x,y)=xy-x^3$ we easily check that $b_i=b_j$
for all $i,j$ due to symmetry. Hence, we should definitely look at \emph{nonlinear} kernels.

As above, we can carry out calculations for the more general 
case of the polynomial kernel (M2). We obtain 
\benn
\int_\cX (1+xz)^p \sqrt{\frac{\alpha_i}{\pi}}\txte^{-\alpha_i(x-b_i)^2}~\txtd x
\approx \int_\R (1+xz)^p \sqrt{\frac{\alpha_i}{\pi}}\txte^{-\alpha_i(x-b_i)^2}~\txtd x=:I_p.
\eenn
The last integral can be evaluated for any $p\in \N$ and we just give a few examples
\beann
&I_1= 1+b_iz,\quad I_2=\frac{2 \alpha_i (b_i z+1)^2+z^2}{2 \alpha_i},\quad 
I_3= \frac{(b_i z+1) \left(2 \alpha_i (b_i z+1)^2+3 z^2\right)}{2 \alpha_i}\\
& I_4= \frac{4 \alpha_i^2 (b_i z+1)^4+12 \alpha_i z^2 (b_i z+1)^2+3 z^4}{4 \alpha_i^2},
\quad 
I_5= \frac{(b_i z+1) \left(4 \alpha_i^2 (b_i z+1)^4+20 \alpha_i z^2 
(b_i z+1)^2+15 z^4\right)}{4 \alpha_i^2}.
\eeann
For $I_1$ we just get a shifted linear kernel, which only depends upon the moving 
average/mean $b_i$, while for all higher ($p\geq 2$) exponents, the MMD also 
depends upon the moving variance $a_i$. We know from the theory of stochastic
fast-slow systems that far away from the critical transition the variance is 
very small, at least in the linearized approximation, so we formally
take the limit $\alpha_i\ra \I$, which yields
\be
\lim_{\alpha_i\ra +\I} I_p=
\int_\R (1+xz)^p \delta_{b_i}(x)~\txtd x=(1+b_iz)^p.
\ee
where $\delta_{b_i}$ denotes the delta-distribution centered at $b_i$. The result 
tells us that the MMD
\be
\lim_{\alpha_i,\alpha_j\ra \I}\mbox{MMD}_{\textnormal{(M2)}}[\cH,P_i, P_j]\approx 
\sup_{\|u\|\leq 1} \langle u,(1+b_i\cdot)^p-(1+b_j\cdot )^p\rangle
\ee
becomes more independent of the (scaled) variance $\alpha_j$ away the critical transition. 
Approaching the critical transition itself, we have to take the limit $\alpha_i\ra 0$, 
in which case $I_p$ becomes
\benn
\lim_{\alpha_i\ra 0}I_p=+\I,\qquad \text{for $p\geq 2$ and $z\neq 0$}.
\eenn
From this result, one might be tempted to conjecture that the MMD could become infinite
as we approach the critical transition. Yet, this is a very subtle point in practice. 
For example, consider $p=2$, then we have to consider the difference
\be
\label{eq:funkeystep}
\frac{2 \alpha_i (b_i z+1)^2+z^2}{2 \alpha_i}-\frac{2 \alpha_j (b_j z+1)^2+z^2}{2 \alpha_j}.
\ee
Let us calculate this expression for the fold bifurcation. We get with 
$\bar{t}_j:=(t_j+t_{j+1})/2$ that
\benn
\label{eq:bjajfold}
\alpha_j=\frac{1}{\sigma_*^2}\sqrt{-\varepsilon \bar{t}_j+y(0)} \qquad \text{and}\qquad  
b_j=\sqrt{-\varepsilon \bar{t}_j+y(0)}.
\eenn
Now let us assume that we just make a small time step for the estimate between the 
two densities so that $\bar{t}_j=\bar{t}_i+\delta$. Plugging this into~\eqref{eq:bjajfold}
and Taylor-expanding yields to first-order in $\delta$ that  
\benn
\alpha_i\approx \alpha_j-\frac{\varepsilon\delta}{2\sigma_*^2\sqrt{y(0)-\bar{t}_j\varepsilon}} 
\qquad \text{and}\qquad  
b_i\approx b_j-\frac{\varepsilon\delta}{2\sqrt{y(0)-\bar{t}_j\varepsilon}}.
\eenn
Note that the last formula contains \emph{three} small parameters: $\delta$, $\varepsilon$, 
$\sigma_*$. This very subtle in practical terms since the relative ratios now matter
crucially as the critical fold transition is approached. Plugging in the approximations
we get that the difference in~\eqref{eq:funkeystep} is approximately equal to 
\benn
\frac{1}{4} \delta  z \varepsilon  \left(\frac{z \left(\frac{\sigma_*^4}{(y(0)-\bar{t}_j\varepsilon
)^{3/2}}-4\right)}{\sigma_*^2}-\frac{4}{\sqrt{y(0)-\bar{t}_j \epsilon }}\right)+\cO(\delta^2)=
:\zeta(z;\delta,\varepsilon,\bar{t}_j).
\eenn
So we conclude that for polynomial kernels (M2) it might be theoretically possible to 
use the MMD as a warning sign since we have
\be
\label{eq:p2main}
\mbox{MMD}_{\textnormal{(M2),fold}}[\cH,P_i, P_j] \approx \sup_{\|u\|\leq 1}
\langle u,\zeta(\cdot;\delta,\varepsilon,\bar{t}_j) \rangle,
\ee
which does diverge for fixed $\delta$ as the fold critical transition is approached when 
$\varepsilon\bar{t}_j\ra y(0)$. However, if $\delta$ is very small (see also 
Section~\ref{sec:models}) then we simply will not see any change in the MMD up to 
a very small neighbourhood of the transition. However, we still see that the
MMD is an excellent binary classifier as the mean between the pre-jump and post-jump
distributions are very different. 

Next, we will use the Gaussian kernel, which is actually characteristic for any moments. 
It is among the most widely used kernels in machine learning. We get that
\benn
I_\sigma:=\int_\R \txte^{\left(-(x-z)^2/\sigma^2\right)}
\sqrt{\frac{\alpha_i}{\pi}}\txte^{-\alpha_i(x-b_i)^2}~\txtd x 
=\sqrt{\frac{\alpha_i}{\alpha_i+\frac{1}{\sigma ^2}}} 
\txte^{-\frac{\alpha_i \mu ^2}{\alpha_i \sigma ^2+1}}.
\eenn
It is useful to consider the two limits, away from the critical transition
for $\alpha_i\ra \I$ and near the transition for $\alpha_i\ra 0$. For the former
we get
\be
\label{eq:limitsGG}
\lim_{\alpha_i\ra +\I} I_\sigma=\txte^{-\frac{(b_i-z) ^2}{\sigma ^2}}.
\ee 
Hence, the MMD satisfies
\be
\label{eq:destr}
\lim_{\alpha_i,\alpha_j\ra \infty}\mbox{MMD}_{\textnormal{(M3)}}[\cH,P_i, P_j]\approx 
\sup_{\|u\|\leq 1}\left\langle u, \left(\txte^{-\frac{(b_i-\cdot)^2}{\sigma ^2}}
-\txte^{-\frac{(b_j-\cdot)^2}{\sigma^2}}\right)\right\rangle
\ee
far away from the transition and the difference in means is crucial. Very
similar calculations as in the polynomial case show that the time scale $\delta$ in
comparing the distribution as well as the sizes of $\varepsilon$ and $\sigma_*^2$
restrict the MMD somewhat as a practical early-warning sign. In summary, we have 
seen that the MMD is expected to be an excellent way to detect 
critical transitions, when the underlying dynamical system is given by a stochastic fast-slow
SODE. The kernel parameters and fast-slow system parameters crucially matter to attenuate 
or decrease certain detection and early-warning scenarios as shown by the (semi-)explicit 
formulas above.

\section{Model Problems and Time Series}
\label{sec:models}

\subsection{An MMD Estimator}
\label{sec:estimator}

We give with a brief description regarding the numerical calculation of the MMD 
based upon the theory in Section~\ref{sec:kernel}. Following up on 
formula~\eqref{eq:MMDdef2}, one can check that given two i.i.d~samples 
$(x_1,\cdots,x_M)$ from $P$ and $(z_1,\cdots,z_M)$ from $Q$, an unbiased estimate 
of MMD is~\cite{Grettonetal} 
\be
\label{eq:MMDest}
\mbox{MMD}_{\textnormal{ub}}^2 := \frac{1}{M(M-1)}\sum_{i \ne j}^M H(\eta_i,\eta_j),
\ee
where $\eta_i := (x_i,z_i)$ and $H(\eta_i,\eta_j) := K(x_i,x_j)+K(z_i,z_j)-
K(x_i,z_j)-K(x_j,z_i)$ and convergence takes place in distribution. When $P\neq Q$, 
it can be shown that, under certain assumptions~\cite{Grettonetal}, 
\benn
\frac{\mbox{MMD}_{\textnormal{ub}}^2-\textnormal{MMD}(\cH,P,Q)}{\sqrt{V_l(P,Q)}} 
\xrightarrow[\text{}]{\text{D}} {\cal N}(0,1), 
\eenn
where $V_M(P,Q)=\cO(M^{-1})$ as $M\ra \I$ and $\cN(0,1)$ is the standard normal distribution. 
When $P=Q$, it can be shown that, under certain assumptions~\cite{Grettonetal},
\benn
M \; \mbox{MMD}_{\textnormal{ub}}^2 \xrightarrow{D} \sum_{\ell = 1}^{\infty} 
\lambda_{\ell} (w_{\ell}^2-2),
\eenn
where $w_{\ell} \sim {\cal N}(0,2)$ i.i.d., $\lambda_i$ are solutions of the 
eigenvalue problem for the centered kernel $\tilde{K}(x_i,x_j) := {K}(x_i,x_j) - 
{\mathbb E}_xK(x_i,x) - {\mathbb E}_x K(x,x_j)+
 {\mathbb E}_{x,x^{\prime}} K(x,x^{\prime})$. In particular, the estimator~\eqref{eq:MMDest}
we are going to use here is known to give a controlled numerical approximation to the
MMD if the sample size $M$ is sufficiently large.  

\subsection{Numerical MMD for a van der Pol Oscillator}
\label{sec:model1}

In this section, we study the fast-slow SODE~\eqref{eq:fs} given by a
van der Pol-type oscillator with
\be
\label{eq:model1}
f=y-\frac{27}{4} x^2(x+1),\quad g=-\frac{1}{2}-x,\quad F=1,
\quad G=1,\quad\sigma_f=0.1,\quad \sigma_g=0.1,\quad \varepsilon=0.01, 
\ee
which was used originally in~\cite{ZhangHallerbergKuehn} to study the reliability of 
early-warning signs for critical transitions.\medskip 

\begin{figure}[htbp]
	\centering
	\begin{overpic}[width=0.75\textwidth]{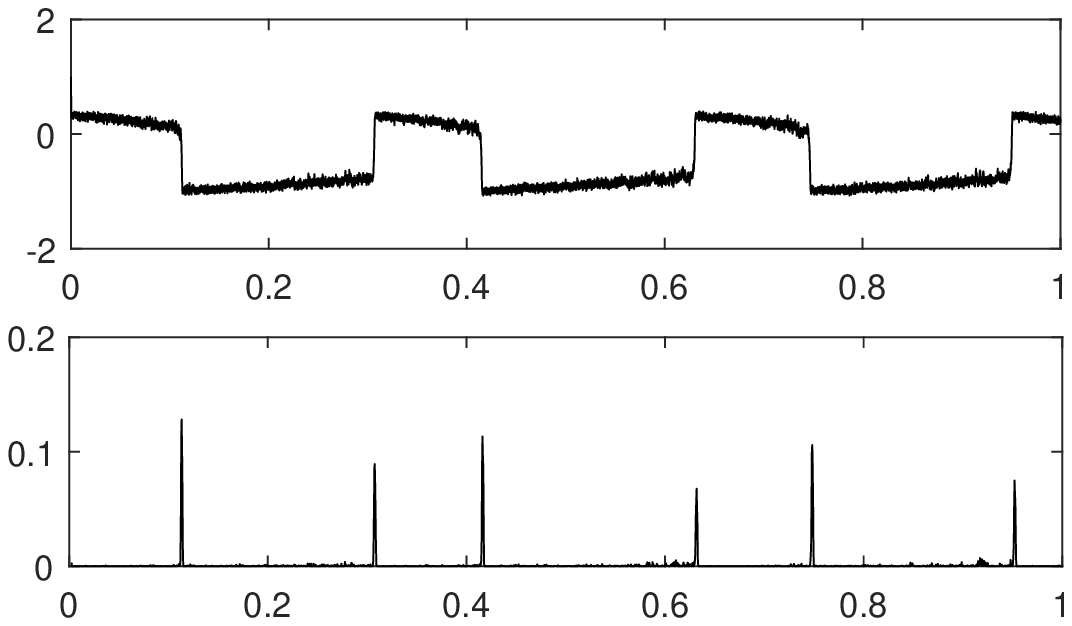}
		\put(-3,15){\rotatebox{90}{$\mbox{MMD}_{\textnormal{ub}}^2$}}
		\put(0,52){$x$}
		\put(53,1){$t$}
		\put(53,31){$t$}
	\end{overpic}
	\caption{\label{fig:01}Numerical computations for the fast-slow SODE~\eqref{eq:fs}
	with detailed model specification~\eqref{eq:model1}. (a) Time series for the fast
	variable $x$ showing the typical (noisy) relaxation oscillations of van der Pol-type
	oscillators. (b) Plot of the squared MMD using the estimator 
	$\mbox{MMD}_{\textnormal{ub}}^2$ for a polynomial kernel with $p=2$.}
\end{figure}

We numerically integrate the 
equations using standard SODE algorithms~\cite{Higham} generating a single time series 
of $5\cdot 10^6$ points, which is shown in Figure~\ref{fig:01}(a), where we normalized
the horizontal time axis to the unit interval $[0,1]$. We analyzed the time
series for the fast variable and computed the MMD~\eqref{eq:MMDest} as shown in 
Figure~\ref{fig:01}(b) for a polynomial kernel as defined in (M2) with $p=2$. Note that 
even if the polynomial kernel defined in (M2) is not characteristic, it still can 
offer some insight on changes of the low order moments. In the 
computation each sliding window consists of $M=1000$ points, within which we estimate 
the MMD in comparison to a window shifted by $200$ points by using~\eqref{eq:MMDest}.\medskip   

\begin{figure}[htbp]
	\centering
	\begin{overpic}[width=0.75\textwidth]{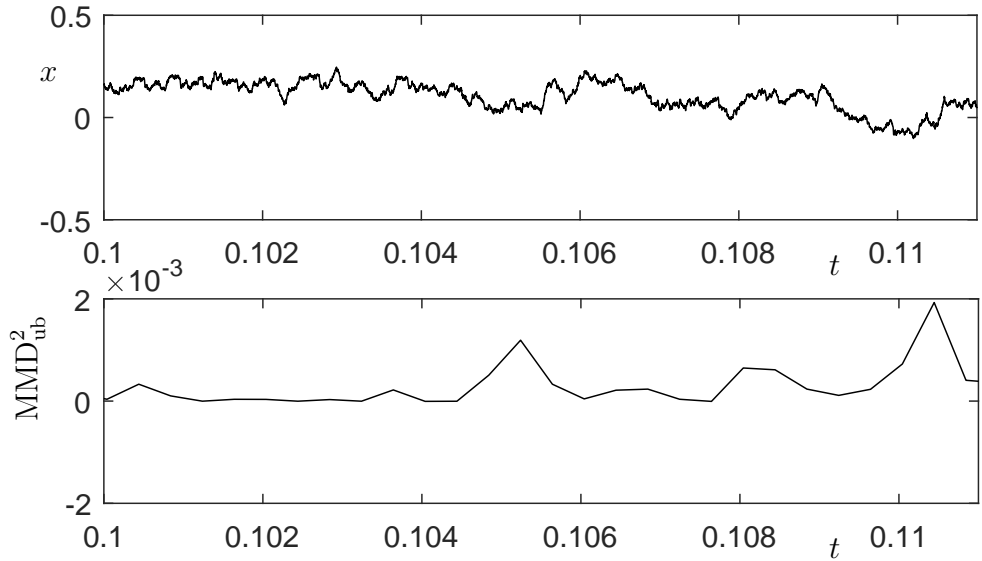}
		\put(-3,15){\rotatebox{90}{$\mbox{MMD}_{\textnormal{ub}}^2$}}
		\put(0,50){$x$}
		\put(81,1){$t$}
		\put(81,30){$t$}
	\end{overpic}
	\caption{\label{fig:02}Same computation as in Figure~\ref{fig:01}. The figure just shows 
	a zoom near the first critical transition.}
\end{figure}

\begin{figure}[htbp]
	\centering
	\begin{overpic}[width=0.795\textwidth]{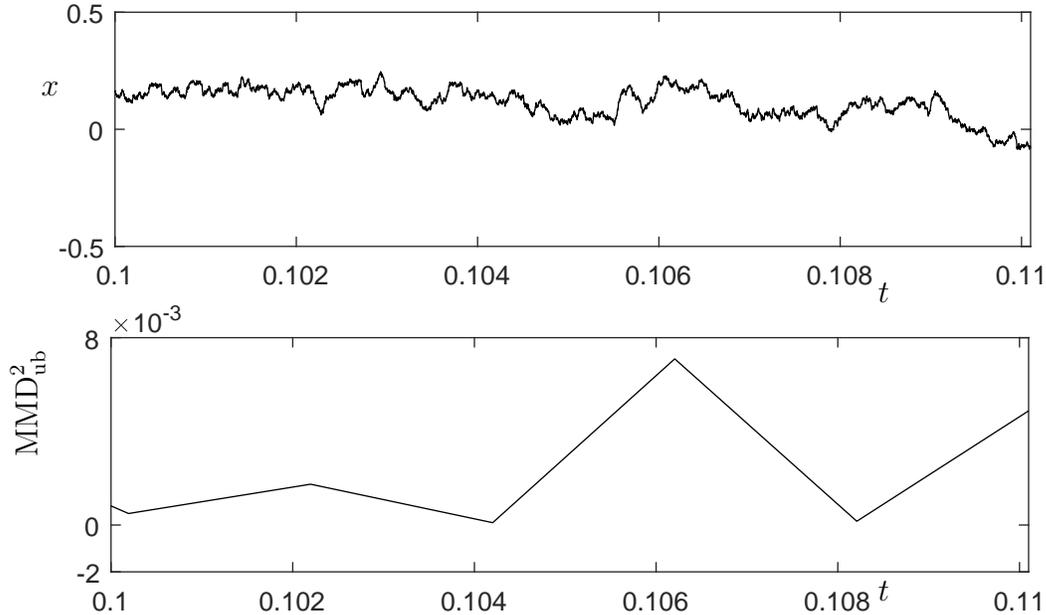}
		\put(-3,15){\rotatebox{90}{$\mbox{MMD}_{\textnormal{ub}}^2$}}
		\put(0,50){$x$}
		\put(81,1){$t$}
		\put(81,30){$t$}
	\end{overpic}
	\caption{\label{fig:04}Same computation as in Figure~\ref{fig:02} except that we now select
	a large shift in the sliding window, instead of $200$ points we take $1000$ points, i.e., 
	non-overlapping windows. The MMD is now larger by about a factor four to five, precisely
	as predicted by the theory from equation~\eqref{eq:p2main} as we have increased essentially
	$\delta$ by a factor five. This means we get a slightly more pronounced warning sign, yet
	it is still probably too small to be practically useful.}
\end{figure}

We observe that the MMD is an excellent binary classifier for the critical transitions
generated at the two fold points of the critical manifold $\cC_0=\{f=0\}$. It is
also easy to check computationally that the results are qualitatively the same for all 
types of polynomial kernels for $p\in\{2,3,4,\ldots\}$. Between the fast jumps at the
fold points, the MMD is very small. Once the jump at the fold is included, then a large 
spike in the MDD is observed as predicted by our theory from Section~\ref{sec:OU}. In 
particular, change-point detection can easily be automated. Near each jump, we can also 
look at a smaller scale before the critical transition. Figure~\ref{fig:02} shows the 
MMD near the first critical transition. There seems to be a small 
tendency to increase before the transition on a much smaller scale as already 
predicted from the theoretical considerations in Section~\ref{sec:OU} as $\delta$
is indeed small in our simulations in Figure~\ref{fig:02}. If we increase the window
shift from $200$ to $1000$ (i.e., non-overlapping but adjacent windows), we 
can indeed see a slightly better early-warning sign as shown in Figure~\ref{fig:04}.
However, the warning sign seems still to hold only very close to the transition so it
might not be practically feasible. Furthermore, if we increase the shift even more,
the choice of sliding window is crucial, which we simply do not know in advance for
a single critical transition. Yet, if we have many transitions, it might be
possible to configure the MMD to always compare the current distribution to a 
learned pre-tipping reference distribution.\medskip

\begin{figure}[htbp]
	\centering
	\begin{overpic}[width=0.75\textwidth]{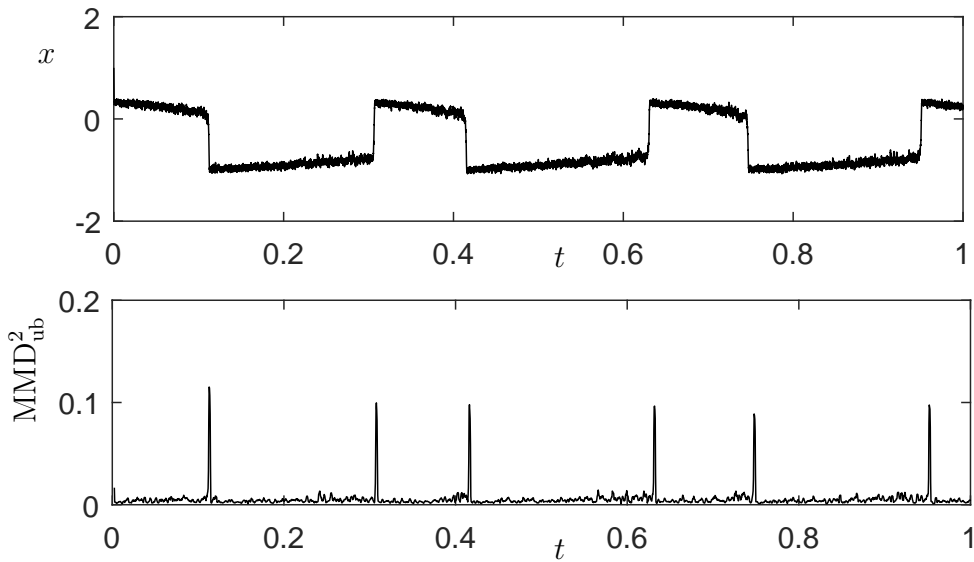}
		\put(-3,15){\rotatebox{90}{$\mbox{MMD}_{\textnormal{ub}}^2$}}
		\put(0,52){$x$}
		\put(53,1){$t$}
		\put(53,31){$t$}
	\end{overpic}
	\caption{\label{fig:03}Numerical computations for the fast-slow SODE~\eqref{eq:fs}
	with detailed model specification~\eqref{eq:model1}. (a) Time series for the fast
	variable $x$ showing the typical (noisy) relaxation oscillations of van der Pol-type
	oscillators. (b) Plot of the squared MMD using the estimator 
	$\mbox{MMD}_{\textnormal{ub}}^2$ for a Gaussian kernel with $\sigma=1$.}
\end{figure}

In Figure~\ref{fig:03}, we show the MMD for the Gaussian kernel from (M3) with parameter
$\sigma=1$. The conclusions are similar to the polynomial kernel that also the Gaussian
kernel performs excellent in detecting the critical transitions, i.e., it is a very
good binary classifier. The only key difference is the larger baseline fluctuations
in the Gaussian case in comparison to the polynomial case in Figure~\ref{fig:01}.



\begin{thebibliography}{10}

\bibitem{Aronszajn}
N.~Aronszajn.
\newblock Theory of reproducing kernels.
\newblock {\em Trans. Amer. Math. Soc.}, 68:337--404, 1950.

\bibitem{AshwinWieczorekVitoloCox}
P.~Ashwin, S.~Wieczorek, R.~Vitolo, and P.~Cox.
\newblock Tipping points in open systems: bifurcation, noise-induced and
  rate-dependent examples in the climate system.
\newblock {\em Phil. Trans. R. Soc. A}, 370:1166--1184, 2012.

\bibitem{BerglundGentz}
N.~Berglund and B.~Gentz.
\newblock {\em Noise-Induced Phenomena in Slow-Fast Dynamical Systems}.
\newblock Springer, 2006.

\bibitem{BoettingerHastings}
C.~Boettinger and A.~Hastings.
\newblock Quantifying limits to detection of early warning for critical
  transitions.
\newblock {\em J. R. Soc. Interface}, 9(75):2527--2539, 2012.

\bibitem{Borgwardtetal}
K.M. Borgwardt, A.~Gretton, M.J. Rasch, H.P. Kriegel, B.~Sch{\"o}lkopf, and
  A.J. Smola.
\newblock Integrating structured biological data by kernel maximum mean
  discrepancy.
\newblock {\em Bioinformatics}, 22(14):e49--e57, 2006.

\bibitem{CarpenterBrock}
S.R. Carpenter and W.A. Brock.
\newblock Rising variance: a leading indicator of ecological transition.
\newblock {\em Ecology Letters}, 9:311--318, 2006.

\bibitem{CuckerSmale}
F.~Cucker and S.~Smale.
\newblock On the mathematical foundations of learning.
\newblock {\em Bull. Amer. Math. Soc.}, 39(1):1--49, 2002.

\bibitem{DitlevsenJohnsen}
P.D. Ditlevsen and S.J. Johnsen.
\newblock Tipping points: early warning and wishful thinking.
\newblock {\em Geophys. Res. Lett.}, 37:19703, 2010.

\bibitem{DumortierRoussarie}
F.~Dumortier and R.~Roussarie.
\newblock {\em Canard Cycles and Center Manifolds}, volume 121 of {\em Memoirs
  Amer. Math. Soc.}
\newblock AMS, 1996.

\bibitem{Fenichel4}
N.~Fenichel.
\newblock Geometric singular perturbation theory for ordinary differential
  equations.
\newblock {\em J. Differential Equat.}, 31:53--98, 1979.

\bibitem{Fryzlewicz}
P.~Fryzlewicz.
\newblock Wild binary segmentation for multiple change-point detection.
\newblock {\em Ann. Stat.}, 42(6):2243--2281, 2014.

\bibitem{Gardiner}
C.~Gardiner.
\newblock {\em Stochastic Methods}.
\newblock Springer, Berlin Heidelberg, Germany, 4th edition, 2009.

\bibitem{Grettonetal}
A.~Gretton, K.M. Borgwardt, M.J. Rasch, B.~Sch{\"o}lkopf, and A.~Smola.
\newblock A kernel two-sample test.
\newblock {\em J. Mach. Learn. Res.}, 13:723--773, 2012.

\bibitem{Hamzi1}
J.~Bouvrie, B.~Hamzi.
\newblock Kernel Methods for the Approximation of Nonlinear Systems.
\newblock {\em SIAM J. Control Optim.}, 55(4), 2460--2492. 2017.

\bibitem{Hamzi2}
B.~Hamzi, T. N.~ AlOtaiby, S.~ AlShebeili, A.~AlAnqary.
\newblock Preliminary Results on a Maximum Mean Discrepancy Approach for Seizure Detection.
\newblock {\em Prof. of the 20th International Conference on Health Informatics and Health Information Technology}, London/UK.

\bibitem{Harchaouietal}
Z.~Harchaoui, F.~Bach, O.~Capp{\'e}, and E.~Moulines.
\newblock Kernel-based methods for hypothesis testing: a unified view.
\newblock {\em IEEE Signal Proc. Mag.}, 30(4):87--97, 2013.

\bibitem{Higham}
D.J. Higham.
\newblock An algorithmic introduction to numerical simulation of stochastic
  differential equations.
\newblock {\em SIAM Review}, 43(3):525--546, 2001.

\bibitem{Jones}
C.K.R.T. Jones.
\newblock Geometric singular perturbation theory.
\newblock In {\em Dynamical Systems (Montecatini Terme, 1994)}, volume 1609 of
  {\em Lect. Notes Math.}, pages 44--118. Springer, 1995.

\bibitem{KruSzm3}
M.~Krupa and P.~Szmolyan.
\newblock Extending geometric singular perturbation theory to nonhyperbolic
  points - fold and canard points in two dimensions.
\newblock {\em SIAM J. Math. Anal.}, 33(2):286--314, 2001.

\bibitem{KuehnCT2}
C.~Kuehn.
\newblock {A mathematical framework for critical transitions: normal forms,
  variance and applications}.
\newblock {\em J. Nonlinear Sci.}, 23(3):457--510, 2013.

\bibitem{KuehnCurse}
C.~Kuehn.
\newblock The curse of instability.
\newblock {\em Complexity}, 20(6):9--14, 2015.

\bibitem{KuehnBook}
C.~Kuehn.
\newblock {\em Multiple Time Scale Dynamics}.
\newblock Springer, 2015.
\newblock 814 pp.

\bibitem{Kuznetsov}
Yu.A. Kuznetsov.
\newblock {\em Elements of Applied Bifurcation Theory}.
\newblock Springer, New York, NY, 3rd edition, 2004.

\bibitem{Liuetal}
S.~Liu, M.~Yamada, N.~Collier, and M.~Sugiyama.
\newblock Change-point detection in time-series data by relative density-ratio
  estimation.
\newblock {\em Neural Networks}, 43:72--83, 2013.

\bibitem{MisRoz}
E.F. Mishchenko and N.Kh. Rozov.
\newblock {\em Differential Equations with Small Parameters and Relaxation
  Oscillations (translated from Russian)}.
\newblock Plenum Press, 1980.

\bibitem{Schefferetal}
M.~Scheffer, J.~Bascompte, W.A. Brock, V.~Brovkhin, S.R. Carpenter, V.~Dakos,
  H.~Held, E.H. van Nes, M.~Rietkerk, and G.~Sugihara.
\newblock Early-warning signals for critical transitions.
\newblock {\em Nature}, 461:53--59, 2009.

\bibitem{SchoelkopfSmola}
B.~Sch{\"o}lkopf and A.J. Smola.
\newblock {\em Learning with Kernels: support vector machines, regularization,
  optimization, and beyond}.
\newblock MIT Press, 2001.

\bibitem{Schoenberg}
I.J. Sch{\"o}nberg.
\newblock On certain metric spaces arising from euclidean spaces by a change of
  metric and their imbedding in {Hilbert} space.
\newblock {\em Ann. Math.}, 38(2):787--793, 1937.

\bibitem{SiegemundVenkatraman}
D.~Siegemund and E.S. Venkatraman.
\newblock Using the generalized likelihood ratio statistic for sequential
  detection of a change-point.
\newblock {\em Ann. Stat.}, 23(1):255--271, 1995.

\bibitem{sri} 
 B.~ K. Sriperumbudur, A. Gretton, K. Fukumizu, B. Schölkopf, G. R. G. Lanckriet
\newblock Hilbert Space Embeddings and Metrics on Probability Measures
\newblock{Journal of Machine Learning Research}, 11, 1517-1561, 2010

\bibitem{Wahba}
G.~Wahba.
\newblock {\em Spline Models for Observational Data}.
\newblock SIAM, 1990.

\bibitem{Wiesenfeld1}
K.~Wiesenfeld.
\newblock Noisy precursors of nonlinear instabilities.
\newblock {\em J. Stat. Phys.}, 38(5):1071--1097, 1985.

\bibitem{Wittenetal}
I.H. Witten, E.~Frank, M.A. Hall, and C.J. Pal.
\newblock {\em Data Mining: Practical machine learning tools and techniquess}.
\newblock Morgan Kaufmann, 2016.

\bibitem{ZhangHallerbergKuehn}
X.~Zhang, S.~Hallerberg, and C.~Kuehn.
\newblock Predictability of critical transitions.
\newblock {\em Phys. Rev. E}, 92:052905, 2015.

\end{thebibliography}
\end{document}